\documentclass{eas}
\usepackage{graphicx}
\def\spose#1{\hbox to 0pt{#1\hss}}
\def\simlt{\mathrel{\spose{\lower 3pt\hbox{$\mathchar"218$}}
     \raise 2.0pt\hbox{$\mathchar"13C$}}}
\def\simgt{\mathrel{\spose{\lower 3pt\hbox{$\mathchar"218$}}
     \raise 2.0pt\hbox{$\mathchar"13E$}}}

\def\eg{{\rm e.g.}}
\begin{document}

\title{Chemodynamical Simulations of Elliptical Galaxies}

\runningtitle{Simulating Elliptical Galaxies}
\author{Brad K. Gibson}\address{University of Central Lancashire,
Centre for Astrophysics, Preston, PR1~2HE, United Kingdom; 
\email{bkgibson@uclan.ac.uk\ \&\ psanchez-blazquez@uclan.ac.uk\ \&\ 
scourty@uclan.ac.uk}}
\author{Patricia S\'anchez-Bl\'azquez}\sameaddress{1}
\author{St\'ephanie Courty}\sameaddress{1}
\author{Daisuke Kawata}\address{Carnegie Observatories, Pasadena, USA;
\email{dkawata@ociw.edu}}
\begin{abstract}
We review recent developments in the field of chemodynamical simulations of
elliptical galaxies, highlighting (in an admittedly biased fashion) 
the work conducted with our cosmological N-body/SPH code {\tt GCD+}.  
We have demonstrated previously the
recovery of several primary {\it integrated}
early-type system scaling
relations (\eg\ colour-magnitude relation, 
L$_{\rm X}$-T$_{\rm X}$-[Fe/H]$_{\rm X}$) 
when employing a phenomenological AGN heating scheme in conjunction
with a self-consistent treatment of star formation, 
supernovae feedback, radiative cooling, chemical enrichment, and stellar/X-ray
population synthesis.  Here we emphasise characteristics derived
from the full {\it spatial} information contained within the simulated 
dataset, including stellar and coronal morphologies, metallicity distribution
functions, and abundance gradients.
\end{abstract}
\maketitle
\section{Introduction}
Our understanding of the formation and evolution of elliptical
galaxies is undergoing something of a renaissance.  The comfortable
classical hierarchical merging scenario in which massive ellipticals
form later than their less massive counterparts, faces challenges 
when confronted with empirical evidence for the existence of old,
metal-rich, proto-ellipticals at redshifts $z$$\approx$2$\rightarrow$3.
The concepts of 'downsizing' (in which stars in more massive galaxies
form earlier and on a shorter timescale than in less massive systems)
and 'dry mergers' (dissipationless merging of stellar systems, without
associated star formation) have proven to be popular enhancements to the
conventional picture, although both face challenges in the face 
of the constraints imposed by known correlations in optical
(\eg\ colour-magnitude, luminosity-metallicity, Fundamental Plane)
and X-ray (\eg\ luminosity-temperature-metallicity) properties.

While the semi-analytical methodology remains powerful when exploring 
both monolithic (\eg\ Gibson 1997) and merger-driven (\eg\ Pipino \etal\
2006) frameworks for galaxy formation, the computational approach
has progressed to the stage where fully coupled multi-wavelength
spectro-chemical + dynamical simulations of ellipticals are now feasible.
For example, the
origin of optical scaling relations and abundance gradients has
been explored recently by Kobayashi (2004), 
within the idealised context of semi-cosmological initial conditions, while
the power of fully cosmological hydrodynamic simulations, which neglect both
supernovae and AGN feedback, has been 
demonstrated by Dom\'inguez-Tenreiro \etal\ (2006) and 
Naab \etal\ (2006).  Recent cosmological simulations which include 
treatments of feedback include those of Meza \etal\ 
(2003),\footnote{Neglecting AGN feedback.} Sommer-Larsen \etal\ 
(2003),\footnote{Neglecting AGN feedback, Type Ia supernovae, and 
metallic line cooling.} and Kawata \& Gibson (2005, and references
therein).

Kawata \& Gibson (2005) studied the effect of AGN heating on the chemodynamical
evolution of ellipticals, including the impact on both the 
integrated optical (\eg\ colour-magnitude relation) and X-ray
(\eg\ luminosity-temperature-metallicity relations) properties.  Driven
by central convergent gas flows, the self-regulating nature of 
AGN heating within the Kawata \& Gibson simulation leads to a stable hot
corona and the suppression of late-time star formation.  The cosmological
elliptical simulated was shown to be consistent with these observed
integrated properties.

\section{New Results}
Instead of repeating the discussion of Kawata \& Gibson (2005), in what 
follows we report on our recent work in assessing the success (and
failure) of our current suite of simulations in recovering various 
spatially-resolved optical and X-ray observables.  This report
is very much a work-in-progress, but should provide the reader with 
an overview of the direction in which we are heading.

\subsection{Morphology}

Our reference simulation here corresponds to Model~2 of 
Kawata \& Gibson (2005), a mildly triaxial, slow oblate rotator, whose
central line-of-sight velocity dispersion $\sigma$ and rotation velocity 
V$_{\rm rot}$ match those of, for example, NGC~3379 ($\sim$50~km/s and
$\sim$250~km/s, respectively).\footnote{Admittedly though, the simulated 
galaxy's effective radius
is $\sim$11~kpc, while that of NGC~3379 is $\sim$2~kpc.}  
Making use of its simulated optical 
isophotes, we measured the $m$=4 deviations from 
perfect ellipses (the so-called $a_4$ parameter).  Viewed at high-inclination
the galaxy can be characterised as possessing disky ($a_4$$>$0) isophotes
over the range 0.5$\simlt$$R/R_{eff}$$\simlt$1.5, and boxy
($a_4$$<$0) isophotes beyond $R/R_{eff}$$\approx$2 (see Fig~1), 
consistent with that found by Meza \etal\ (2003; Fig~11; 
50$^\circ$$\simlt$$\theta$$\simlt$70$^\circ$) for their cosmological 
elliptical galaxy simulation, and the disky isophotes observed 
within the effective radius of NGC~3379.

\begin{figure}
\begin{center}
\includegraphics[scale=0.3]{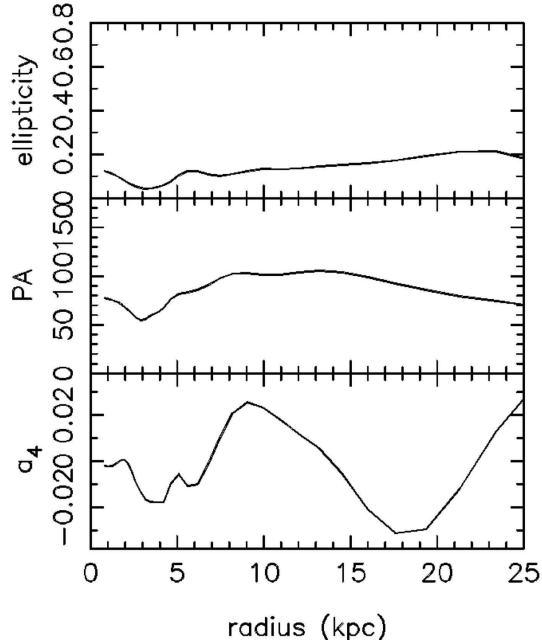}
\caption{Ellipticity, position angle, and $m$=4 deviations from perfect
ellipses (top to bottom panels, respectivey) for Model~2 of Kawata
\& Gibson (2005), as viewed in projection at high-inclination.}
\end{center}
\end{figure}

Diehl \& Statler (2006) have recently suggested that the hot coronal gas
within $R/R_{eff}$$<$2
and the underlying stellar components of ellipticals appear to have broadly
aligned position angles, but the ellipticities are not well-correlated,
as might be expected in hydrostatic equilibrium.  We attempted to 
confront our simulations with this Chandra-inspired
empirical evidence by fitting ellipses
to the X-ray contour maps of our reference simulation; unfortunately,
insufficient resolution compromised our ability to do so.  It
is clear though that the next-generation of simulations will be 
ideally suited for this problem (Gottl\"ober \etal\ 2006).


\subsection{Gradients}
In the upper (left) panel of 
Fig~2, we show the light-weighted iron (blue) and oxygen (red) abundance
gradients for the stellar component of Kawata \& Gibson's (2005) 
Model~2; the bottom (left) panel shows the corresponding
abundance ratio ([O/Fe]) gradient.  Near the effective radius, the
gradients in [Fe/H] and [O/Fe] are d$<$[Fe/H]$>$/d(log~$r$)$\approx$$-$0.3
and d$<$[O/Fe]$>$/d(log~$r$)$\approx$$+$0.1, respectively.  The agreement
with the gradients derived by S\'anchez-Bl\'azquez et~al. 
(2006)\footnote{Also, for the first time, measured at $R/R_{eff}$$\simgt$1.}
for NGC~3379 (d$<$[Fe/H]$>$/d(log~$r$)$\approx$$-$0.30 and
d$<$[O/Fe]$>$/d(log~$r$)$\approx$$+$0.04, respectively) is re-assuring
(and admittedly, perhaps a coincidence) and consistent with the similarity
found between the morphology of NGC~3379 and our reference model
(\S~2.1).  Our simulated elliptical is 
consistent with the predictions of "outside-in" scenarios, albeit with
a somewhat shallower [O/Fe] gradient than the default model of 
Pipino \etal\ (2006).\footnote{Which has a steeper gradient:
d$<$[O/Fe]$>$/d(log~$r$)$\approx$$+$0.3.}

\begin{figure}
\begin{center}
\includegraphics[scale=0.25]{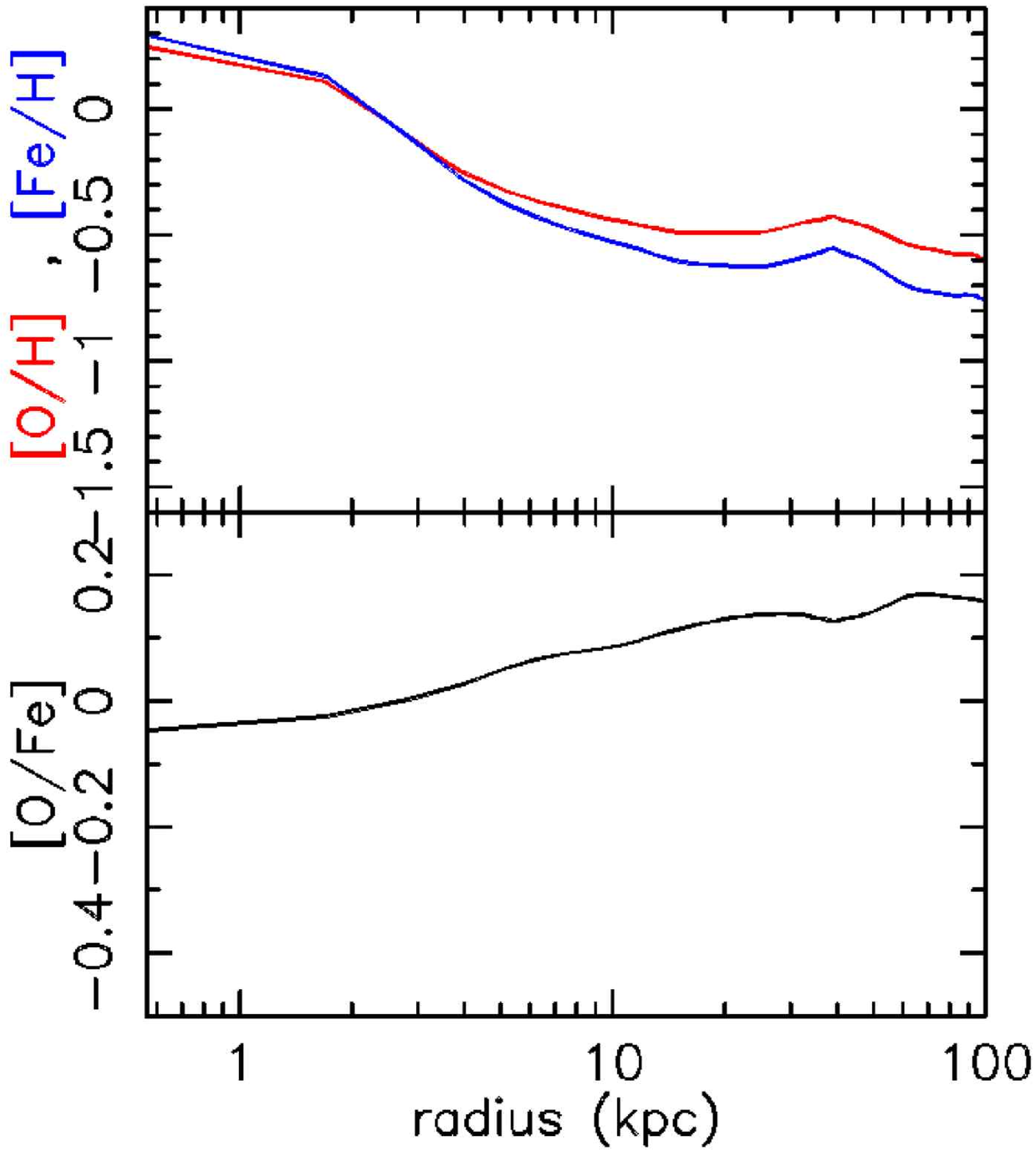}
\includegraphics[scale=0.25]{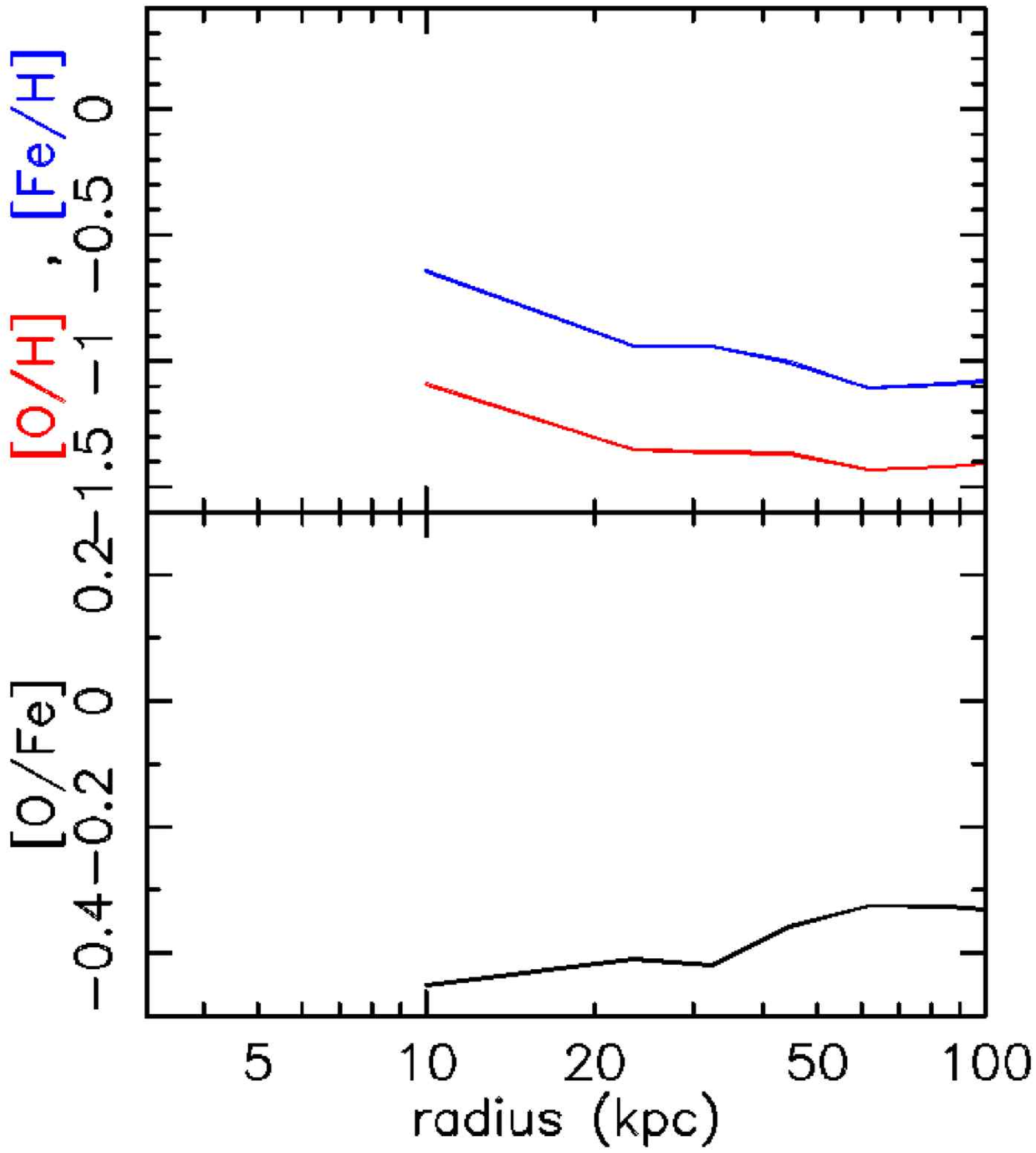}
\caption{{\it Left:} Stellar abundance (upper) and abundance ratio (lower)
gradients for Model~2 of Kawata \& Gibson (2005). {\it Right:} X-ray coronal
abundances (upper) and abundance ratio (lower) gradients for the same
simulation.}
\end{center}
\end{figure}

Over the range 1$\simlt$$R/R_{eff}$$\simlt$5, 
the abundance (and abundance ratio)
gradients in the hot corona of our simulated elliptical
are d$<$[Fe/H]$_{\rm X}$$>$/d(log~$r$)$\approx$$-$0.6 (a factor of two
steeper than the iron gradient in the stellar component at 
$R/R_{eff}$$\simlt$1) and d$<$[O/Fe]$_{\rm X}$$>$/d(log~$r$)$\approx$$+$0.1
(which matches that of the stellar component at $R/R_{eff}$$\simlt$1,
albeit, obviously offset by a factor of five in magnitude - see above),
respectively.  This abundance ratio is consistent with 
a scenario whereby $\sim$70\% of the iron associated with the hot corona
originated from Type~Ia supernovae (Gibson \etal\ 1997).

\subsection{Metallicity Distribution Functions \& Abundance Gradients}
Using spatially-resolved metallicity distribution functions (MDFs) from HST,
Harris \& Harris (2002) infer a metallicity gradient for the stellar
population of NGC~5128 of d($<$[m/H]$>$)/d(log~r$_e$)$\approx$$-$0.5~dex/log~r 
(for 1.4$\simlt$r$_e$$\simlt$4); the FWHM dispersion of these empirical
MDFs is $\sim$1~dex (see Fig~3).  At the same effective radius, our 
simulated elliptical possesses a flat gradient in [m/H] (again, see
Fig~3), matching such an inferred steep empirical
slope only in the inner region of the galaxy (over the range
0.2$\simgt$r$_e$$\simgt$0.5 - recall Fig~2).  
This should not be surprising given that the
simulation in question possesses gradients more broadly consistent
with NGC~3379 (\S~2.2) than NGC~5128, but for illustrative purposes here
the comparison remains of interest.  Of perhaps more immediate interest
though is the fact that the dispersion in the simulated MDFs 
is $\sim$1.5~dex (FWHM), a factor of 2-3 broader than the empirical MDFs.
Whether this apparent discrepancy is a manifestation of a "G-dwarf" problem, 
contamination of the inner field MDF by a secondary bulge component, or
partially driven by observational selection effects, remains unclear; 
what cannot be disputed though is the potential for future MDF observations
to constrain elliptical galaxy formation scenario (Pipino \etal\ 2006).

\begin{figure}
\begin{center}
\includegraphics[scale=0.5]{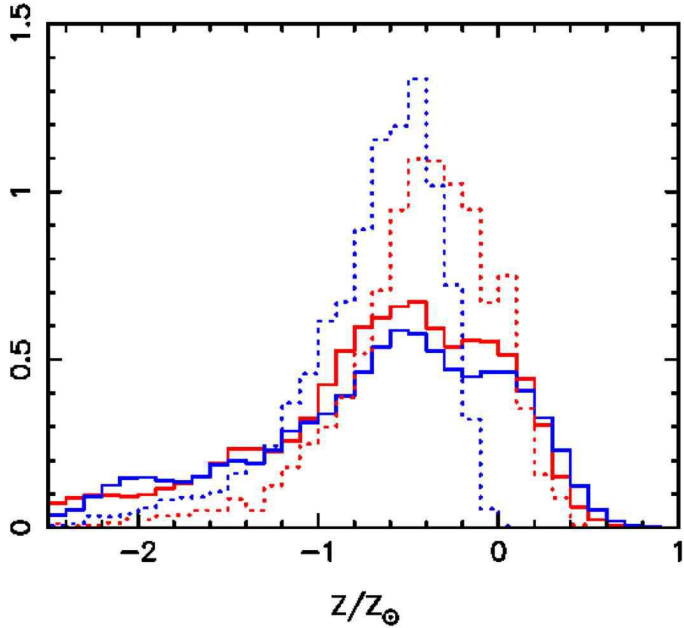}
\caption{Simulated (solid: Model~2 - Kawata \& Gibson 2005) 
and empirical (dotted: NGC~5128 - Harris \& Harris 2002) MDFs for the 
inner (red) and outer (blue) fields described in \S~2.2.}
\end{center}
\end{figure}

\section{Future Directions}
While our past and ongoing work has proven successful across a range
of testables, it is perhaps more interesting to emphasise here several
of the unsolved problems, challenges, and potential pitfalls, remaining.
For example, the observed abundances in the hot coronal gas of 
ellipticals are (roughly speaking) solar (for Fe, Mg, Si, S, and Ar), 
half-solar (O and Ne), and three times solar (Ni), while in the 
hot intracluster medium of galaxy clusters it is 0.3$\times$ solar for
Fe, twice solar for Si, half-solar for S, solar for Ni, and $<$0.1$\times$
solar for Ar and Ca (with significant cluster-to-cluster variations).  
No combination of Type~Ia and Type~II supernovae can be made
consistent with these data, implynig the need for some additional 
metal source(s).  That said,
one remains contrained by empirical supernova rates now derived
out to $z$$\approx$1, while
Pop~III stars and hypernovae do not appear to be a panacea in this regard.
Is the fact that our simulated MDF is $\sim$0.5~dex broader than the 
empirical MDF derived from NGC~5128 of concern?  Is there a serious
G-dwarf problem?  How robust are our X-ray luminosity-weighted 
abundance and abundance ratio gradients at $\sim$3~$R_{eff}$?  Is the 
line-of-sight velocity dispersion skewness $h_3$ correlated with the
sense of direction of the rotation velocity (Meza \etal\ 2003)?
Are the optical and X-ray isophotes correlated in position angle and/or
ellipticity?  What impact might mass-dependent star formation and 
AGN feedback efficiencies play?  Single simulations, such as that
described here or in Meza \etal\ (2003), have proven enticing, but
what is now needed is a significant statistical sample of cosmological
elliptical galaxy simulations before we can properly 
address the nature of these empirical trends.


\end{document}